\documentclass[aps,prb,twocolumn,showpacs,preprintnumbers,superscriptaddress]{revtex4}
\usepackage{amssymb}

\usepackage{graphicx}
\usepackage{dcolumn}
\usepackage{bm}
\usepackage{color}

\begin{document}

\preprint{PREPRINT (\today)}
\title{Diamagnetism, Nernst signal, and finite size effects in superconductors above the transition temperature $T_{c}$}
\author{T.~Schneider}
\email{toni.schneider@swissonline.ch}
\affiliation{Physik-Institut der Universit\"{a}t Z\"{u}rich, Winterthurerstrasse 190, CH-8057 Z\"{u}rich, Switzerland}
\author{S.~Weyeneth}
\affiliation{Physik-Institut der Universit\"{a}t Z\"{u}rich, Winterthurerstrasse 190, CH-8057 Z\"{u}rich, Switzerland}

\begin{abstract}
Various superconductors, including cuprate superconductors, exhibit peculiar features above the transition temperature $T_{c}$. In particular the observation of a large diamagnetism and Nernst signal $N$ in a wide temperature window above $T_{c}$ attracted considerable attention. Noting that this temperature window exceeds the fluctuation dominated regime drastically and that in these materials the spatial extent of homogeneity is limited, we explore the relevance of the zero dimensional (0D)-model, neglecting thermal fluctuations. It is shown that both, the full 0D-model as well as its Gaussian approximation, mimic the essential features of the isothermal magnetization curves $m_{\mathrm{d}}\left(H\right)$ in Pb nanoparticles and various cuprates remarkably well. This analysis also provides estimates for the spatial extent of the homogeneous domains giving rise to a smeared transition in zero magnetic field. The resulting estimates for the amplitude of the in-plane correlation length exhibit a doping dependence reflecting the flow to the quantum phase transition in the underdoped limit. Furthermore it is shown that the isothermal Nernst signal of a superconducting Nb$_{0.15}$Si$_{0.85}$ film, treated as $N\propto -m_{\rm d}$, is fully consistent with this scenario. Accordingly, the observed diamagnetism above $T_c$ in Pb nanoparticles, in the cuprates La$_{1.91}$Sr$_{0.09}$CuO$_{4}$ and BiSr$_2$Ca$_2$CuO$_{8-\delta}$, as well as the Nernst signal in Nb$_{0.15}$Si$_{0.85}$ films, are all in excellent agreement with the scaling properties emerging from the 0D-model, giving a universal perspective on the interplay between diamagnetism, Nernst signal, correlation length, and the limited spatial extent of homogeneity. Our analysis also provides evidence that singlet Cooper pairs subjected to orbital pair breaking in a 0D system are the main source of the observed diamagnetism and Nernst signal in an extended temperature window above $T_{c}$. \end{abstract}

\maketitle

%
%*******************************************

%now the abstract***************************

%*******************************************
%~\\

%
%\narrowtext
%

%*******************************************

\section{Introduction}

\begin{figure}[htb]
%\centering
\includegraphics[width=1.0\linewidth]{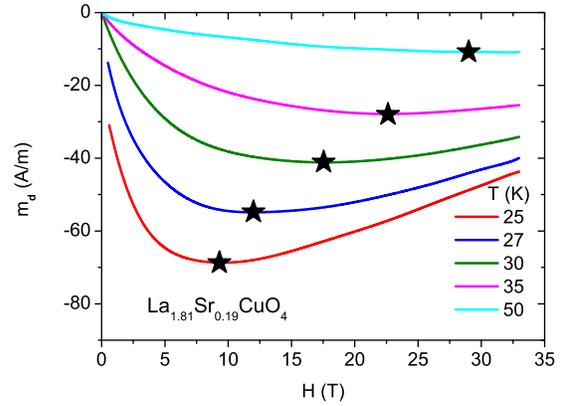}  \vspace{-0.5cm}
\caption{(color online) The solid lines show isotherms $m_{\mathrm{d}}$ \textit{vs}. $H$ at various temperatures above $T_{c}\simeq 23$~K for La$_{1.91}$Sr$_{0.09}$CuO$_{4}$ taken from Li \textit{et al}.\protect\cite{ong}. The symbol $\bigstar$ indicates the minimum in the curves for the presented isotherms.}
\label{fig1}
\end{figure}

The detection of Cooper pairs above the superconducting transition
temperature $T_{c}$ has a long history, dating back to 1969.\cite%
{tinkham} It implies that the average of the order parameter squared, $%
\langle \left\vert \psi \right\vert ^{2}\rangle $, does not
vanish above $T_{c}$, either due to thermal fluctuations or the limited
effective spatial extent of the system. While the regime where thermal
fluctuations dominate is reasonably well understood in terms of the scaling
theory of critical phenomena subjected to finite size effects,\cite%
{ffh,tsh,kosh,hofer,book,parks,tsjs,larkin} novel features have been
observed considerably outside the fluctuation dominated regime.
Recently, Li \textit{et al}.\cite{li,ong} have compiled the results of an extended experimental study of the isothermal magnetization of several families of cuprate superconductors over a rather broad range of temperatures and magnetic fields. From these, they infer that above the transition temperature $T_{c}$, the isothermal diamagnetic contribution to the magnetization $m_{\mathrm{d}}$ decreases initially with increasing magnetic field $H$, applied parallel to the $c$-axis, consistent with $m_{\mathrm{d}}=-\chi _{\mathrm{d}}H$, where $\chi _{\mathrm{d}}$ is the diamagnetic susceptibility. However, as $H$ increases $m_{\mathrm{d}}$ tends to a minimum at $H_{\mathrm{m}}$ and in excess of this characteristic field the magnetization increases and appears to approach zero, as shown in Fig.~\ref{fig1} for La$_{1.91}$Sr$_{0.09}$CuO$_{4}$, showing data taken from Li \textit{et al}.\cite{ong}

Closely related behavior was reported earlier for oriented powder samples of underdoped Y$_{1-x}$Ca$_{x}$Ba$_{2}$Cu$_{3}$O$_{y}$,\cite{lascial} for Pb nanoparticles with particle size larger than the correlation length,\cite{bernardi} for MgB$_{2}$,\cite{lasicalmgb2} and for polycrystalline SmBa$_{2}$Cu$_{3-y}$Al$_{y}$O$_{6+\delta }$.\cite{bernardi2} The magnetization was either inferred from torque magnetometry,\cite{li,ong} or probed by SQUID magnetometery.\cite{lascial,bernardi,bernardi2} However, the torque measurements reveal a temperature dependent paramagnetic background $m_{\mathrm{p}}\left(T,H\right)$. The diamagnetic contribution to the magnetization is then derived from\cite{ong}
\begin{equation}
m_{\mathrm{d}}\left(T,H\right)=m_{\mathrm{exp}}\left(T,H\right)-m_{\mathrm{p}}\left(T,H\right),  \label{eq01}
\end{equation}
where
\begin{equation}
m_{\mathrm{p}}\left(T,H\right)\simeq\left(a+bT\right)H,~~~~~a>>bT.
\label{02}
\end{equation}
Although this subtraction leads to uncertainties in the high field limit, where the diamagnetic signal becomes small, it appears unlikely that the main feature, the occurrence of the minimum in $m_{\mathrm{d}}\left(T,H\right)$ at fixed temperature $T>T_{c}$, is an artifact of this subtraction. Although the crossover from the initial linear ($m_{\mathrm{d}}=-\chi _{\mathrm{d}}H$) to nonlinear behavior is an expected feature of thermal fluctuations in homogeneous two (2D) and three (3D) dimensional superconductors, the occurrence of the minimum cannot be explained invoking these scenarios.\cite{ffh,tsh,kosh,hofer,book,parks} On the other hand, there is considerable evidence that cuprate and amorphous conventional superconductors are homogeneous over a limited spatial domain only. \cite{tsjs,pan,lang,iguchi,tsbled,loram,tscastro,tstool} In this case, the growth of the correlation lengths is limited by approaching the transition temperature $T_{c}$ because it cannot exceed the respective extent of the homogenous domains. Within a two dimensional superconductor, consisting of a stack of superconducting layers with insulating spacing sheets in between, the adoption of this scenario where the magnetization stems from homogeneous domains with limited extent only would result in an effective 0D-superconductor. As a consequence, the isothermal magnetization curves would exhibit a minimum, reminiscent to the observation in nanoparticles.\cite{bernardi}

Related behavior was also observed in the isothermal Nernst signal of superconducting and amorphous Nb$_{0.15}$Si$_{0.85}$ films above $T_{c}$.\cite{pourret1,pourret,pourret2} In this system, the Nernst signal due to normal quasiparticles is particularly low. This allows to probe the contribution associated with superconductivity without the subtraction of the background due to the normal quasiparticles.\cite{pourret1,pourret,pourret2} At low magnetic field, the Nernst signal $N=\nu H$ increases linearly with field, where $\nu$ is the Nernst coefficient. Upon increasing the magnetic field, $N$ deviates from this linear field dependence, reaches a maximum at $H_{\rm m}$ and decreases afterwards. In analogy to the position of the minimum in the magnetization, the maximum shifts to higher fields with increasing temperature. This behavior confirms the evidence that the isothermal Nernst signal is proportional to the magnetization in terms of $N\propto -m_{\rm d}$.\cite{caroli,huse,wang,wang2}

Here we review the properties of the isothermal magnetization curves of a 0D-superconductor, neglecting thermal fluctuations, and explore the consistency with experimental magnetization data of Pb nanoparticles,\cite{bernardi} bulk La$_{1.91}$Sr$_{0.09}$CuO$_{4}$,\cite{ong} and BiSr$_2$CaCu$_2$O$_{8+\delta}$ (Bi2212) with $T_{c}\simeq45$~K and $T_{c}\simeq85$~K.\cite{ong} To explore wether this scenario also accounts for the Nernst signal $N$ in terms of the relation $N\propto -m_{\rm d}$, we consider the data for superconducting Nb$_{0.15}$Si$_{0.85}$ films taken above $T_{c}$.\cite{pourret1,pourret,pourret2} In Sec.~II we sketch the theoretical background including the properties of the 0D-model. The neglect of thermal fluctuations implies that the model is applicable outside the critical regime only, that is
sufficiently above $T_{c}$, the regime where the experimental data of La$_{1.91}$Sr$_{0.09}$CuO$_{4}$ and Bi2212 was taken. Invoking quantum scaling the doping dependence of the minimum in the isothermal magnetization curves is also addressed. In Sec.~III we present the analysis of the data based on the 0D-model, neglecting thermal fluctuations. The remarkable agreement with the measured isothermal magnetization curves, achieved for reasonable values of the model parameters, suggest that the occurrence of the minimum is attributable to a finite extent of the homogeneous domains. The doping dependence of the Bi2212 data is also consistent with the flow to a quantum phase transition in the underdoped limit. Furthermore it is shown that the profile of the isothermal Nernst signal of the superconducting Nb$_{0.15}$Si$_{0.85}$ film, treated as $N\propto -m_{\rm d}$, is fully consistent with the 0D-model. Accordingly, singlet Cooper pairs subjected to orbital pair breaking in a 0D system are the main source of the observed diamagnetism and Nernst signal in an extended temperature window above $T_{c}$.  Finally we show that the
0D-model provides for a variety of conventional and hole doped
superconductors a universal perspective on the interplay between
diamagnetism, Nernst signal, correlation length and the limited spatial
extent of homogeneity. We close with a brief summary and some discussion.

\section{Theoretical background}

\label{seq:Theoretical_background}

The fluctuation contribution to the free energy per unit volume of a homogeneous and anisotropic type II superconductor scales above $T_{c}$ as\cite{ffh,tsh,hofer,book,parks}
\begin{equation}
f=\frac{k_{\mathrm{B}}T}{\xi _{x}\xi _{y}\xi _{z}}G\left( \frac{\xi _{x}\xi_{y}}{L_{H_{z}}^{2}}\right),~~~~L_{H_{z}}^{2}=\frac{\Phi _{0}}{H_{z}},
\label{eq1}
\end{equation}
where $G\left( z\right)$ is a scaling function of its argument and $L_{H_{z}} $ is the magnetic field induced limiting length giving rise to a finite size effect.\cite{tsjpcm} We assume that the magnetic field is applied along the $z$-axis. $\xi _{x,y,z}$ denote the correlation length along the respective axis in zero field. In the limit $\xi _{x}\xi_{y}>>L_{H_{z}}^{2}=\Phi_{0}/H_{z}$, attainable for sufficiently high fields, this expression reduces to
\begin{equation}
f\propto \frac{k_{\mathrm{B}}T}{L_{H_{z}}^{2}\xi _{z}}=\frac{k_{\mathrm{B}}TH_{z}}{\Phi_{0}\xi _{z}},
\label{eq2}
\end{equation}
because the zero field correlation lengths $\xi _{x}\xi _{y}$\ cannot grow beyond $L_{H_{z}}^{2}$. In this limit the magnetization $m_{\mathrm{d}}=-\partial f/\partial H_{z}$ tends to
\begin{equation}
\frac{m_{\mathrm{d}}}{T}\propto -\frac{k_{\mathrm{B}}}{\Phi _{0}\xi _{z}}.
\label{eq3}
\end{equation}
On the other hand, in the opposed limit $\xi _{x}\xi_{y}<<L_{H_{z}}^{2}=\Phi_{0}/H_{z}$ the scaling function adopts the limiting behavior, $G\left(z\right)\propto z^{2}$, to recover $m_{\mathrm{d}}$ $=-\chi _{\mathrm{d}}H$. In this case we obtain
\begin{equation}
\frac{m_{\mathrm{d}}}{T}\propto -\frac{2k_{\mathrm{B}}\xi _{x}\xi _{y}}{\Phi
_{0}\xi _{z}}H_{z}.
\label{eq4}
\end{equation}
Accordingly, in both the 3D and 2D case, where $\xi _{z}=d$ and $d$ denotes the thickness of the superconducting sheets, the magnetization saturates at sufficiently high fields due to the magnetic field induced finite size effect, reducing the effective dimensionality $D$ of the system from $D$ to $D-2$.\cite{tsjpcm,palee} In this limit the system corresponds in $D=3$ to independent superconducting cylinders of radius $L_{H_{z}}\propto \left(\Phi_{0}/H_{z}\right) ^{1/2}$ and height $\xi _{z}$ and in $D=2$ with height $\xi _{z}=d$. Detailed calculations in $D=2$ reveal that the crossover from the low to the high field limit occurs monotonically and according to that there is no minimum.\cite{kosh} 

So far we considered homogeneous systems only. In practice any real and highly anisotropic type II superconductor is homogenous within \textit{e.g.} a cylinder of radius $R$ and height $d$. Concentrating on temperatures sufficiently above $T_{c}$ where thermal fluctuations in the phase and amplitude of the order parameter can be neglected, we are left with a 0D-system with an order parameter $\psi$ which does not depend on the space variables. The temperature and magnetic field dependence follows then from
the Ginzburg-Landau (GL) model for a 0D system as treated by Shmidt.\cite{shmidt,larkin} The partition function in this case reads
\begin{equation}
Z=\int d\text{Re}(\psi)d\text{Im}(\psi)\text{Exp}\left( -f\left[ \psi \right]\right),
\label{eq5}
\end{equation}
with the GL free energy functional
\begin{eqnarray}
f\left[\psi\right] &=& \frac{V}{k_{\mathrm{B}}T}\Bigg[r_{0}\left(\ln \left(\frac{T}{T_{c}}\right) +\left( \frac{2\pi }{\Phi _{0}}\right)^{2}\xi_{0}^{2}\left\langle \mathbf{A}^{2}\right\rangle \right) \left\vert\psi\right\vert^{2}  \nonumber \\
&+&\frac{u_{0}}{2}\left\vert \psi\right\vert^{4}\Bigg],
\label{eq6}
\end{eqnarray}
and
\begin{equation}
t=T/T_{c0}-1,\text{ }\xi _{0}^{2}=\frac{\hbar ^{2}}{2mr_{0}},\text{ }\xi^{2}=\xi _{0}^{2}t^{-1}.
\label{eq7}
\end{equation}
$\mathbf{A}$ is the vector potential and $\xi$ the correlation length with amplitude $\xi _{0}$. Setting
\begin{eqnarray}
\psi^{2} &=&\left\vert z\right\vert ^{2}\frac{k_{\mathrm{B}}T}{V},\nonumber \\
a&=&r_{0}\left(\ln\left( \frac{T}{T_{c}}\right) +\left( \frac{2\pi }{\Phi _{0}}\right) ^{2}\xi_{0}^{2}\left\langle \mathbf{A}^{2}\right\rangle \right),\nonumber \\
u &=&u_{0}\left( \frac{k_{\mathrm{B}}T}{V}\right) ,
\label{eq8}
\end{eqnarray}
assuming $a>0$ and $u>0$ we obtain for the partition function the expression
\begin{equation}
Z=\frac{\pi ^{3/2}k_{\mathrm{B}}T}{V\sqrt{2u}}\text{Exp}\left( \frac{a^{2}}{2u}\right) \text{Erfc}\left( \frac{a}{\sqrt{2u}}\right) .
\label{eq9}
\end{equation}
Erfc$\left( z\right) $ is the complementary error function. The magnetization per unit volume follows then from the free energy $F=-k_{\mathrm{B}}T\ln (Z)$ in terms of
\begin{equation}
m_{\mathrm{d}}=-\frac{1}{V}\frac{dF}{dH}=\frac{k_{\mathrm{B}}T}{V}\cdot\frac{1}{Z}\frac{dZ}{dH},
\label{eq9p5}
\end{equation}
yielding
\begin{equation}
m_{\mathrm{d}}=\frac{k_{\mathrm{B}}T}{V}\left( \frac{a}{u}\frac{da}{dH}+\frac{d}{dH}\ln\left[\text{Erfc}\left( \frac{a}{\sqrt{2u}}\right) \right]\right) .
\label{eq10}
\end{equation}
Using the gauge $\mathbf{A}=(0,H_{z}x,0)$ we obtain for a cylindrical homogenous domain with radius $R$, height $d$ and a spherical domain with radius $r$
\begin{equation}
\left\langle \mathbf{A}^{2}\right\rangle =\frac{H^{2}}{V}\int x^{2}dV=a_{4}H^{2},~~~H=H_{z},
\label{eq11}
\end{equation}
where
\begin{equation}
a_{4}=\left\{\begin{array}{l}R^{2}/4\text{,}~~~V=\pi R^{2}d \\
r^{2}/5,~~~V=4\pi r^{3}/3
\end{array}
\right.
\label{eq12}
\end{equation}
The magnetization expression (\ref{eq10}) can then be rewritten as
\begin{equation}
m_{\mathrm{d}}=a_{3}\left( \frac{4a_{1}H}{a_{2}^{2}}x+\frac{d}{dH}\ln\left[\text{Erfc}\left(x\right)\right]\right),
\label{eq13}
\end{equation}
or
\begin{eqnarray}
m_{\mathrm{d}} &=& \frac{2a_{3}a_{1}}{a_{2}}\left( \frac{x}{a_{1}}-\ln\left( \frac{T}{T_{c}}\right) \right) ^{1/2}\cdot  \nonumber \\
& & \Bigg(2x+\frac{d}{dx}\ln \left[\text{Erfc}\left(x\right)\right]\Bigg),
\label{eq14}
\end{eqnarray}
where
\begin{eqnarray}
x &=&a_{1}\left( \ln \left( \frac{T}{T_{c}}\right) +\left( \frac{H}{a_{2}}\right) ^{2}\right) ,  \nonumber \\
a_{1} &=&\frac{r_{0}V^{1/2}}{\sqrt{2u_{0}k_{\mathrm{B}}T}},~a_{2}=\frac{\Phi_{0}}{2\pi \xi _{0}a_{4}^{1/2}},~a_{3}=\frac{k_{\mathrm{B}}T}{V}.
\label{eq15}
\end{eqnarray}
In terms of the variable $x$, requiring the values of $a_{1}$ and $a_{2}$, Eq.~(\ref{eq13}) adopts the simple scaling form
\begin{equation}
\frac{m_{\mathrm{d}}}{H}=\frac{2a_{3}a_{1}}{a_{2}^{2}}f\left(x\right),~f\left(x\right) =2x+\frac{d}{dx}\ln \left[ \text{Erfc}\left( x\right)\right],
\label{eq15a}
\end{equation}
with the limiting behavior 
\begin{eqnarray}
\left. f\left( x\right)\right\vert_{x\rightarrow \infty }&=&-1/x,\\\nonumber
\left. f\left( x\right)\right\vert_{x\rightarrow 0}&=&-2/\sqrt{\pi }+\left( 2-4/\pi \right) x. 
\end{eqnarray}
In the limit $x\rightarrow 0$ $m_{\mathrm{d}}$ reduces then to
\begin{equation}
m_{\mathrm{d}}=-\frac{4a_{3}a_{1}H}{a_{2}^{2}\sqrt{\pi }},
\label{eq16}
\end{equation}
consistent with $m_{\mathrm{d}}$ $=-\chi _{\mathrm{d}}H$. Contrariwise, for $x=\infty $ it approaches
\begin{equation}
m_{\mathrm{d}}=-\frac{2a_{3}H}{a_{2}^{2}}\left( \ln \left( \frac{T}{T_{c}}\right)+\left( \frac{H}{a_{2}}\right) ^{2}\right) ^{-1}.
\label{eq17}
\end{equation}
Accordingly, the isothermal magnetization curves adopt for $T>T_{c}$ a minimum between the low and high field limits. This characteristic behavior also appears in Fig.~\ref{fig1}. More specific the minimum at
\begin{equation}
x_{\mathrm{m}}\left( t\right) =a_{1}\left( \ln \left( \frac{T}{T_{c}}\right)+\left(\frac{H_{\mathrm{m}}\left( t\right) }{a_{2}}\right) ^{2}\right)
\label{eq17a}
\end{equation}
follows from
\begin{equation}
\frac{dm_{\mathrm{d}}}{dH}=0,
\label{eq18}
\end{equation}
yielding for sufficiently large $a_{1}$ the solution
\begin{equation}
x_{\mathrm{m}}\left( T\right) =x_{\mathrm{m}}\left( T_{c}\right) +2a_{1}\ln \left(\frac{T}{T_{c}}\right),
\label{eq19}
\end{equation}
where
\begin{equation}
x_{\mathrm{m}}\left( T_{c}\right) =a_{1}\left( \frac{H_{\mathrm{m}}\left(T_{c}\right) }{a_{2}}\right) ^{2}\simeq1.02634.
\label{eq19a}
\end{equation}
Together with Eq. (\ref{eq17a}) and sufficiently large $a_{1}$ we obtain for the magnetic field $H_{\mathrm{m}}\left( T\right)$, where the isothermal magnetization curves adopt a minimum, the relation
\begin{equation}
H_{\mathrm{m}}\left( T\right) =a_{2}\left( \frac{1.02634}{a_{1}}+\ln \left(\frac{T}{T_{c}}\right) \right) ^{1/2}.
\label{eq19b}
\end{equation}
Accordingly, $H_{\mathrm{m}}\left( T\right) $ does not vanish at $T_{c}$ and the temperature dependent part is related with $a_{2}$ in Eq.~(\ref{eq15}) to the amplitude of the correlation length. This differs from the Gaussian approximation,
valid for $u_{0}=0$ ($a_{1}=\infty $). In this limit Eq.~(\ref{eq17}), rewritten in the form
\begin{equation}
m_{\mathrm{d}}=-2a_{3}\frac{H}{H_{\mathrm{m}}^{2}\left( T\right) }\left(1+\left( \frac{H}{H_{\mathrm{m}}\left( T\right) }\right) ^{2}\right) ^{-1},
\label{eq20}
\end{equation}
applies. Here $m_{\mathrm{d}}\left( H,T\right) /T$ adopts at fixed temperature a minimum at
\begin{eqnarray}
H_{\mathrm{m}}\left( T\right)&=&H_{\mathrm{m0}}\text{ln}^{1/2}\left(\frac{T}{T_{c}}\right),\nonumber\\
H_{\mathrm{m0}}&=&a_{2}=\frac{\Phi _{0}}{2\pi \xi_{0}a_{4}^{1/2}},
\label{eq21}
\end{eqnarray}
between the low and high field behavior. Furthermore, in the cylindrical case is the ratio $\xi _{0}^{2}/d$ according to Eqs.~(\ref{eq12}) and (\ref{eq21}) given by
\begin{equation}
\frac{\xi _{0}^{2}}{d}=\frac{\Phi _{0}^{2}}{\pi k_{\mathrm{B}}T}\frac{a_{3}}{a_{2}^{2}}.
\label{eq21a}
\end{equation}
The Gaussian expression Eq.~(\ref{eq20}) for the magnetization also implies that there is no particular depairing field. Indeed, the magnetization  vanishes as $m_{d}=-2a_{3}/H$.

As aforementioned, the applicability of the Gaussian approximation requires that
\begin{equation}
\frac{r_{0}V^{1/2}}{\sqrt{2u_{0}k_{\mathrm{B}}T}}\left( \ln\left(\frac{T}{T_{c}}\right)+\left(\frac{H}{a_{2}}\right) ^{2}\right) \rightarrow\infty
\label{eq21c}
\end{equation}
is very large. According to this, in systems with non-negligible quartic term $u_{0}$ it fails close to $T_{c}$ in the low field limit. Considering highly anisotropic cuprates, such as La$_{2-x}$Sr$_{x}$CuO$_{4}$ with $x=0.09 $ and Bi2212, this corresponds to the critical regime where a 3D-xy to 2D-xy crossover occurs and phase fluctuations dominate.\cite{wts} However, in this regime and for sufficiently large $R$ even the full model fails because fluctuations are neglected. In the light of these considerations it is not unexpected that the Gaussian version of the model mimics the essential features of the field dependence of the magnetization shown in Fig. \ref{fig1} well, namely the occurrence of the minimum between the low and high field behavior. The same qualitative agreement also emerges from the magnetization data of La$_{2-x}$Sr$_{x}$CuO$_{4}$ with $x=0.06$, \cite{li} $0.055$,\cite{li}, Bi2212 with $T_{c}\approx85$~K, underdoped Bi2212 with $T_{c}\approx 45$~K,\cite{ong} optimally and overdoped Bi2201La$_{y}$ with $T_{c}\simeq30$~K and $T_{c}\simeq 20$~K,\cite{ong}, and Pb particles.\cite{bernardi} To substantiate this qualitative agreement we explore in Sec.~III the consistency of the measured isothermal magnetization curves of Pb,\cite{bernardi} La$_{2-x}$Sr$_{x}$CuO$_{4}$ with $x=0.09$ and Bi2212\cite{ong} with the outlined scenario for a zero dimensional system. Because the partition function (\ref{eq9}) requires that $a>0$ and with that according to Eqs. (\ref{eq8}) and (\ref{eq12})
\begin{equation}
\ln \left( \frac{T}{T_{c}}\right) +\frac{\pi ^{2}}{\Phi _{0}^{2}}\xi_{0}^{2}R^{2}H^{2}>0,
\label{eq21b}
\end{equation}
our analysis is essentially restricted to temperatures $T>T_{c}$.

One also expects that the diamagnetic contribution to the magnetization exhibits a characteristic doping dependence. Indeed, the phase transition line of La$_{2-x}$Sr$_{x}$CuO$_{4}$ is well described by the empirical relation
\begin{eqnarray}
T_{c}\left( x\right) &=&T_{\mathrm{cm}}\left( 1-2\left( \frac{x}{x_{\mathrm{m}}}-1\right)^{2}\right)  \nonumber \\
&=&\frac{2T_{\mathrm{cm}}}{x_{\mathrm{m}}^{2}}\left( x-x_{\rm u}\right) \left(x_{\rm o}-x\right) ,
\label{eq22}
\end{eqnarray}
due to Presland \textit{et al}.\cite{presland} At $T_{c}=0$ the systems are expected to undergo a quantum phase transition. Here the amplitude of the correlation length $\xi _{0}$ diverges in a homogeneous system as\cite{book}
\begin{equation}
\xi _{0}\propto \delta ^{-\overline{\nu }},
 \label{eq23}
\end{equation}
while $T_{c}$ scales according to
\begin{equation}
T_{c}\propto \delta ^{\overline{z\nu }}.
\label{eq24}
\end{equation}
$\delta$ denotes the tuning parameter of the quantum phase transition with dynamic critical exponent $z$ and correlation length exponent $\overline{\nu}$. Combining Eqs.~(\ref{eq23}) and (\ref{eq24}), we obtain
\begin{equation}
\xi _{0}\propto T_{c}^{-1/z},
\label{eq25}
\end{equation}
expected to apply in both, the underdoped ($x=x_{\rm u}$) and overdoped ($x=x_{\rm o}$) limits. Noting that $\xi _{0}$ enters $H_{\mathrm{m0}}=a_{2}=\Phi_{0}/\left( \pi \xi _{0}R\right) $ [Eqs. (\ref{eq12}) and (\ref{eq21})], the magnetic field $H_{\mathrm{m}}=H_{\mathrm{m0}}\left(\ln\left(T/T_{c}\right)\right)^{1/2}$, where the isothermal magnetization curves exhibit a minimum, the approach to the underdoped or overdoped limit should be observable. However, this behavior may be masked by means of the doping dependence of $R$, the radius of the homogeneous cylindrical domains.

\section{Data Analysis}

\label{seq:Datal}

In this section we explore the consistency of isothermal magnetization and Nernst signal data with the predictions of the full 0D-model and its Gaussian version. We concentrate on the magnetization data of Pb nanoparticles,\cite{bernardi} La$_{1.91}$Sr$_{0.09}$CuO$_{4}$,\cite{ong} and bulk BiSr$_2$CaCu$_2$O$_{8+\delta}$ (Bi2212) with $T_{c}\simeq85$~K (slightly underdoped) and $T_{c}\simeq45$~K (heavily underdoped),\cite{ong} and on Nernst signal data of a Nb$_{0.85}$Si$_{0.85}$ film.\cite{pourret1,pourret,pourret2}

\subsection{Pb nanoparticles}

\begin{figure}[htb]
%\centering
\includegraphics[width=1.0\linewidth]{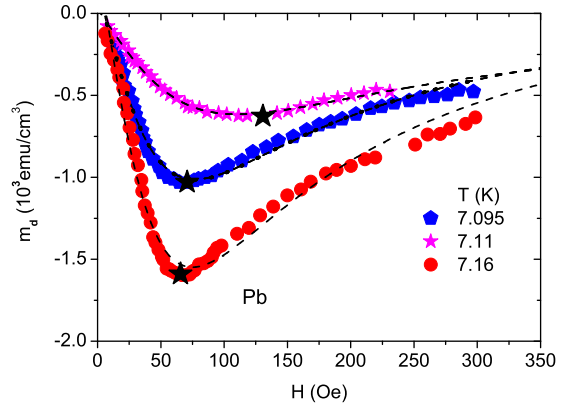}  \vspace{-0.5cm}
\caption{(color online) Diamagnetic magnetization isotherms $m_{\mathrm{d}}$ \textit{vs}. $H$ for Pb nanoparticles with average radius $r=375$~\AA\ taken from Benardi \textit{et al}.\protect\cite{bernardi} The dashed lines are fits to the Gaussian approximation Eq.~(\protect\ref{eq20}) with the parameters listed in Table \protect\ref{Table1}. The dotted line at $T=7.11$~K is a fit to Eq.~(\protect\ref{eq13}) with the parameters listed in Table \protect\ref{Table2} with $T_{c}=7.09$~K. The stars indicate the respective minima of the experimental data at $H_{\mathrm{m}}$.}
\label{fig2}
\end{figure}

The diamagnetism in Pb nanoparticles with average diameters ranging from $150$ to $750$~\AA , sizes for which effects of finite-level spacing should be negligible, has been studied by Benardi \textit{et al}.\cite{bernardi} By means of high-field resolution superconducting quantum interference device (SQUID) measurements, isothermal magnetization curves were obtained for $T\gtrsim T_{c}$. Fig.~\ref{fig2} shows isothermal diamagnetic magnetization curves of the sample containing spherical nanoparticles with average radius $r=375$~\AA . For comparison we included fits to the Gaussian approximation Eq.~(\ref{eq20}) with the parameters listed in Table \ref{Table1} and at $T=7.11$~K to the $0$D-model [Eq. (\ref{eq13})] yielding the parameters
\begin{eqnarray}
a_{1} &=&1303.7, \nonumber \\
a_{2} &=& H_{\mathrm{m0}}=1407.46~\text{Oe},  \nonumber \\
a_{3} &=&0.09~\text{emu Oe/cm}^{3},\nonumber \\
a_{5}&=& 0.00015~\text{emu/cm}^{3},
\label{eq25a}
\end{eqnarray}
with $T_{c}=7.09$~K in terms of the dotted line. To account for a temperature dependent background contribution we added to Eqs.~(\ref{eq13}) and (\ref{eq20}) the parameter $a_{5}$, which turns out to be rather small. The agreement of the $0$D-model with the Gaussian approximation at $T=7.11$~K also reveals that for $a_{1}\approx 1300$ the limit $a_{1}\rightarrow\infty$ is nearly attained. Indeed, the fit parameters $a_{2}$, $a_{3}$ and $a_{5}$ as obtained from the 0D-model [Eq. (\ref{eq25a})] and the Gaussian counterpart (see Table \ref{Table1})\ coincide nearly at $T=7.11$~K.

\begin{table*}[t!]
\caption[~]{ Fit parameters entering the Gaussian approximation [Eq.~(\protect\ref{eq20})] for Pb nanoparticles including the additive background correction $a_{5}$. The amplitude $a_{2}=H_{\mathrm{m0}}$ is obtained from Eq.~(\protect\ref{eq21}) with $T_{c}=7.09$~K.}
\label{Table1}%
\begin{tabular}{ccccc}
\hline\hline
$T$(K) & $a_{3}$(emu Oe/cm$^{3}$) & $H_{\mathrm{m}}$(Oe) & $a_{5}$(emu/cm$^{3}$) & $a_{2}=H_{\mathrm{m0}}$(Oe) \\ \hline
7.095 & 0.14 & 74.79 & 3.4$\times 10^{-4}$ & 2816.8 \\
7.11 & 0.09 & 76.99 & 1.6$\times 10^{-4}$ & 1450.6 \\
7.16 & 0.07 & 110.29 & 0.3$\times 10^{-4}$ & 1112.7 \\ \hline\hline
\end{tabular}%
\end{table*}

\begin{figure}[htb]
%\centering
\includegraphics[width=1.0\linewidth]{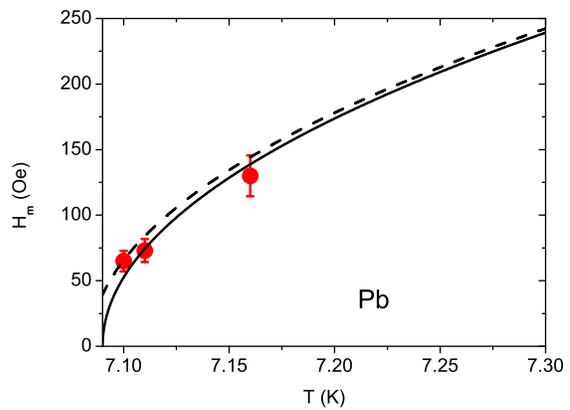}  \vspace{-0.5cm}
\caption{(color online) $H_{\mathrm{m}}$ \textit{vs.} $T$ obtained from the magnetization data of the Pb nanoparticles. The dots mark the respective $H_{\mathrm{m}}$'s of the experimental data shown in Fig.~\protect\ref{fig2} with $\Delta H_{\mathrm{m}}/H_{\mathrm{m}}=0.12$. The solid line is Eq. (\protect\ref{eq21}) with $a_{2}=1400$~Oe and $T_{c}=7.09$~K, and the dashed one Eq.~(\protect\ref{eq19b}) with the parameters listed in Eq.~(\protect\ref{eq25a}).}
\label{fig3}
\end{figure}

On the other hand, considering $H_{\mathrm{m}}$ \textit{vs}. $T$, from Eqs.~(\ref{eq19b}) and (\ref{eq25a}) one expects that this agreement does not hold sufficiently close to $T_{c}$, because $H_{\mathrm{m}}\left(T_{c}\right) $ does not vanish in the full 0D-model for any $a_{1}>0$ [Eq.~(\ref{eq19b})]. In Fig. \ref{fig3} we observe that this behavior is well confirmed. Nevertheless we observe that the Gaussian approximation describes $H_{\mathrm{m}}(T)$ rather well except very close to $T_{c}$. Contrariwise, in this regime thermal fluctuations are no longer negligible and even the 0D-model is not applicable. In this view it is gratifying that the Gaussian approximation describes $H_{\mathrm{m}}(T)$ for sufficiently large $a_{1}$ and away from $T_{c}$ rather well.

Next we turn to the parameter $a_{3}=k_{\mathrm{B}}T/V$. In this context it should be recognized that in the resulting $V$ the packing density of the nanoparticles (spheres) is not taken into account. Indeed, the packing density $\eta $ is the fraction of a volume filled by spheres. Noting that $\eta $ varies from $0.055$ for the loosest possible to $0.7405$ for cubic close packing, it becomes clear that $V$ is not simply related to the radius of the nanoparticles [$V=\left( 4\pi /3\right) r^{3}$]. For $r=375$~{\AA } corresponding to $2.21\times 10^{-16}$~cm$^{3}$ and $V=k_{\rm B}T/a_{3}\simeq10.9\times 10^{-15}$~cm$^{3}$ ($a_{3}=0.09$~emuOe/cm$^3$ and $T=7.11$~K) we obtain $\eta\simeq0.02$, revealing that in the sample considered here the packing density of the nanoparticles is worse than the loosest one. Given the uncertainty in the actual packing density we invoke for the amplitude of the correlation length the estimate $\xi _{0}\simeq 1000$~\AA ,\cite{kerchner} yielding with Eqs. (\ref{eq12}), (\ref{eq15}) and $a_{2}=H_{\mathrm{m0}}=1450$~Oe
\begin{equation}
r=\frac{\Phi _{0}\sqrt{5}}{2\pi \xi _{0}a_{2}}\simeq 508~\text{\AA },
\label{eq26}
\end{equation}
in comparison with $r=375$~\AA, estimated from AFM images of Pb nanoparticles onto a mica substrate.\cite{bernardi}

Even though the Hartree approximation works well for sufficiently high fields and away from $T_{c}$, it should be kept in mind that it fails inevitably in the zero field limit. Here the quartic term in the GL-functional is essential to remove the divergence of the correlation
length $\xi $ at $T_{c}$. Indeed, $\xi $ cannot grow beyond $r$.

\subsection{La$_{1.91}$Sr$_{0.09}$CuO$_{4}$}

\begin{figure}[htb]
%\centering
\includegraphics[width=1.0\linewidth]{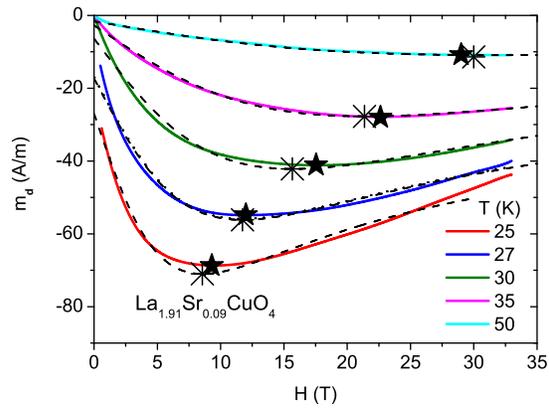}  \vspace{-0.5cm}
\caption{(color online) Isothermal magnetization curves $m_{\mathrm{d}}$ \textit{vs}. $H$ at various temperatures above $T_{c}\simeq 23$~K of La$_{1.91}$Sr$_{0.09}$CuO$_{4}$ taken from Li \textit{et al}.\protect\cite{ong} $\bigstar $ indicate the minima in the experimental and $\divideontimes $ of the Gaussian curves at the respective fixed temperatures. The dashed curves are fits to the Gaussian approximation [Eq.~(\protect\ref{eq20})] with the parameters listed in Table \protect\ref{Table2} and the dotted one to the 0D-model [Eq.~(\protect\ref{eq13})] with the parameters given in Eq.~(\protect\ref{eq26a}). In both models we included the additive background correction $a_{5}$.}
\label{fig4}
\end{figure}

A glance at the isothermal magnetization curves shown in Figs.~\ref{fig1} and \ref{fig2} uncover, surprisingly enough, the same characteristic behavior. Indeed, $m_{\mathrm{d}}$ decreases initially with increasing magnetic field, consistent with $m_{\mathrm{d}}=-\chi _{\mathrm{d}}H$, where $\chi _{\mathrm{d}}$ is the diamagnetic susceptibility. However, as $H$ increases $m_{\mathrm{d}}$ tends to a minimum at $H_{\mathrm{m}}$ and in excess of this characteristic field the magnetization increases and appears to approach zero. Noting that $1$~A/m=$10^{-3}$~emu/cm$^{3}$, the most striking difference concerns the magnitude of $H_{\mathrm{m}}$ which depends on the amplitude of the correlation length in terms of $a_{2}$ in Eq.~(\ref{eq19b}). As $\xi_{0}$ in La$_{1.91}$Sr$_{0.09}$CuO$_{4}$ is around $25$~\AA, \cite{kohout} compared to $1000$~\AA\ in Pb, the difference in $H_{\mathrm{m}}$ is not attributable to $\xi _{0}$ only but points to a substantial large value of $a_{4}^{1/2}=r/\sqrt{5}$ in Pb in comparison with $a_{4}^{1/2}=R/2$ [Eq.~(\ref{eq12})] in La$_{2-x}$Sr$_{x}$CuO$_{4}$, where $r$ is the radius of the nanoparticles, while $R$ is the radius of the homogenous cylindrical domains in the cuprate. To explore these analogies and differences between the isothermal magnetization curves of Pb and La$_{2-x}$Sr$_{x}$CuO$_{4}$ with $x=0.09$ quantitatively, we analyzed the data shown in Fig.~\ref{fig1} on the basis of the Gaussian model [Eq. (\ref{eq20})] yielding the parameters listed in Table \ref{Table2}. For comparison we included in Fig.~\ref{fig4} a fit to the 0D-model [Eq.~(\ref{eq13})] yielding at $T=27$~K the parameters
\begin{eqnarray}
a_{1} &=&285.33, \nonumber\\
a_{2} &=& 29.27~\text{T}, \nonumber\\
a_{3} &=&455.67~\text{AT/m}, \nonumber\\
a_{5}& =& -17.30~\text{A/m}
\label{eq26a}
\end{eqnarray}
where $a_{5}$ is the additive background correction. Though the parameter $a_{1}$ is considerably smaller than its Pb counterpart [Eq.~(\ref{eq25a})] we observe that the fit parameters $a_{2}$, $a_{3}$ and $a_{5}$ as obtained from the 0D-model [Eq.~(\ref{eq25a})] and the Gaussian counterpart (see Table \ref{Table2}) nearly coincide at $T=27$~K. As the Gaussian approximation requires Eq.~(\ref{eq21c}) to be fulfilled, this agreement is attributable to the fact that the temperatures considered here are considerably above $T_{c}\simeq 23$~K.

\begin{table*}[t!]
\caption[~]{Fit parameters entering Eq. (\protect\ref{eq20}) for La$_{1.91}$Sr$_{0.09}$CuO$_{4}$ including the additive background correction $a_{5}$. The amplitude $a_{2}=H_{\mathrm{m0}}$ is obtained from Eq.~(\protect\ref{eq21}) with $T_{c}=23$~K.}
\label{Table2}%
\begin{tabular}{ccccc}
\hline\hline
$T$~(K) & $a_{3}$~(AT/m) & $H_{\mathrm{m}}\left( T\right) $~(T) & $a_{5}$~(A/m)
& $a_{2}=H_{\mathrm{m0}}$(T) \\ \hline
25 & 376.66 & 8.56 & -26.99 & 29.64 \\
27 & 455.69 & 11.73 & -17.29 & 29.29 \\
30 & 561.41 & 15.67 & -6.34 & 30.4 \\
35 & 533.59 & 21.36 & -2.73 & 32.96 \\
50 & 294.95 & 30.01 & -1.63 & 34.04 \\ \hline\hline
\end{tabular}%
\end{table*}

Given the estimate $a_{2}=31$~T we obtain with Eqs.~(\ref{eq12}) and (\ref{eq15})
\begin{equation}
R\xi _{0}=\frac{\Phi _{0}}{\pi a_{2}}\simeq 2.13\times 10^{3}~\text{\AA }^{2},
\label{eq27}
\end{equation}
in comparison with $r\xi _{0}\simeq 5.1\times 10^{5}$~\AA $^{2}$ in Pb corresponding to $a_{2}=940$~Oe. This difference is responsible for the large amplitude of $a_{2}$ of $H_{\mathrm{m}}\left( T\right) $ [Eq.~(\ref{eq19b}] in La$_{1.91}$Sr$_{0.09}$CuO$_{4}$. To estimate the radius $R$ of the homogeneous domains we invoke for the amplitude of the in-plane correlation length the estimate $\xi _{0}\approx 25$~\AA \cite{kohout} yielding with Eq. (\ref{eq27})
\begin{equation}
R=\frac{\Phi _{0}}{\pi \xi _{0}a_{2}}\simeq 85~\text{\AA ,}
\label{eq28}
\end{equation}
which is comparable to the amplitude of the in-plane correlation length. Noting that 1~A/m=$10$~erg/(cm$^{3}$T) we obtain for $d$ the estimate
\begin{equation}
d=\frac{V}{\pi R^{2}}=36~\text{\AA },
\label{eq28a}
\end{equation}
using $T=27$~K and $a_{3}=455.69$~AT/m, $V=\pi R^{2}d=8.2\times 10^{-19}$~cm$^{3}$ and $R=85$~\AA . To check the reliability of these estimates, based on the amplitude $\xi _{0}\approx 25$~\AA,\cite{kohout} we invoke Eq.~(\ref{eq21a}), yielding with $T=27$~K, $\xi _{0}=25$~\AA\ and $d=36$~\AA the estimate $a_{3}/a_{2}^2=0.49$~A/mT, in reasonable agreement with $a_{3}/a_{2}^2=0.53$~A/mT, resulting from the fit listed in Table \ref{Table2}.

\begin{figure}[htb]
%\centering
\includegraphics[width=1.0\linewidth]{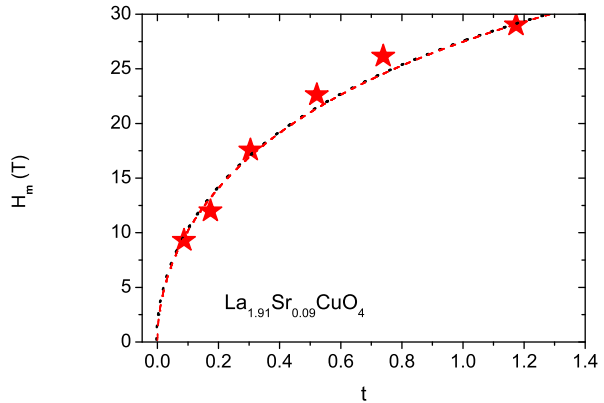}  \vspace{-0.5cm}
\caption{(color online) $H_{\mathrm{m}}$ \textit{vs.} $t=T_c/T-1$ for La$_{1.91}$Sr$_{0.09}$CuO$_{4}$. The dotted line is Eq. (\protect\ref{eq19b}) with $a_{1}=285.33$ and $a_{2}=H_{\mathrm{m0}}=33$~T. The dashed line is the Gaussian approximation with $a_{2}=33$~T [Eq.~(\protect\ref{eq21})]. The stars denote the $H_{\mathrm{m}}$ values derived from the experimental data shown in Fig. \protect\ref{fig4}.}
\label{fig5}
\end{figure}

Given $a_{1}$ and $a_{2}$, the temperature dependence of $H_{\mathrm{m}}$, the magnetic field where the isothermal magnetization curves adopt a minimum, is readily calculated with Eq. (\ref{eq19b}) or Eq. (\ref{eq21}). In Fig. \ref{fig5}, showing the resulting $H_{\mathrm{m}}\left( t\right) $ we observe agreement with the experimental data. Noting that the magnitude of $H_{\mathrm{m}}$ is controlled by the amplitude $H_{\mathrm{m0}}=\Phi _{0}/\left( \pi R\xi _{0}\right)$ and therewith by $R\xi_{0}$, the large difference between $H_{\mathrm{m}}$ of Pb and La$_{1.91}$Sr$_{0.09}$CuO$_{4}$ simply stems from the different $R\xi_{0}$ values, namely $r\xi _{0}\simeq 5.1\times 10^{5}$~\AA $^{2}$ in Pb and $R\xi _{0}\simeq2.13\times 10^{3}$~\AA $^{2}$ in La$_{1.91}$Sr$_{0.09}$CuO$_{4}$. Except for this essential difference we observe a close analogy between the diamagnetic contribution to the isothermal magnetization in Pb nanoparticles and La$_{1.91}$Sr$_{0.09}$CuO$_{4}$. Clearly, this analogy breaks down close to $T_{c}$ and in the zero field limit where thermal fluctuations dominate. Nevertheless, the observed analogy and the agreement with the 0D-scenario, requiring an order parameter $\psi $ which does not depend on the space variables, reveals that in the temperature regime considered here fluctuations can be ignored.

\subsection{Bi2212}

\begin{figure}[htb]
%\centering
\includegraphics[width=1.0\linewidth]{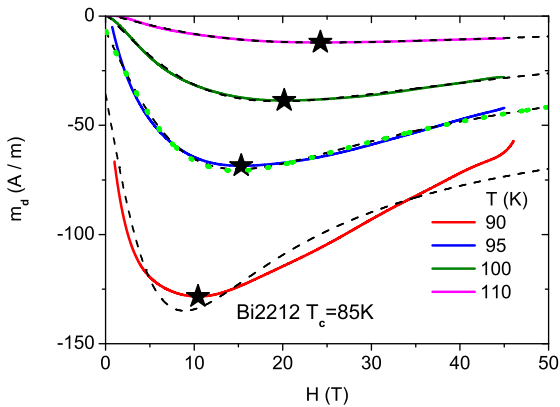}  \vspace{-0.5cm}
\caption{(color online) The solid lines show $m_{\mathrm{d}}$ \textit{vs}. $H$ at various temperatures above $T_{c}\simeq 85$~K of Bi2212 taken from Li \textit{et al}.\protect\cite{ong}. $\bigstar $ indicates the minimum in the isotherms at the respective fixed temperatures. The dashed curves are fits to the Gaussian approximation [Eq.~(\protect\ref{eq20})] with the parameters listed in Table \protect\ref{Table3} and the dotted one to the 0D-model [Eq.~(\protect\ref{eq13})] with the parameters given in Eq.~(\protect\ref{eq28b}). In both models we included the additive background correction $a_{5}$.}
\label{fig6}
\end{figure}

To explore the established analogy between the diamagnetic contribution to the isothermal magnetization in Pb nanoparticles and La$_{2-x}$Sr$_{x}$CuO$_{4}$ with $x=0.09$ further we extend the analysis to the Bi2212 data of Li \textit{et al}.\cite{ong} shown in Fig. \ref{fig6}. We include fits of the Gaussian model [Eq. (\ref{eq20})] yielding the parameters listed in Table \ref{Table3} and a fit to the 0D-model [Eq. (\ref{eq13})] yielding at $T=95$~K,
\begin{eqnarray}
a_1 &=& 311.06, \nonumber\\
a_{2} &=& 44.32~\text{T},  \nonumber \\
a_{3} &=&945.29~\text{AT/m}, \nonumber\\
a_{5} &=& -6.80~\text{A/m.}
\label{eq28b}
\end{eqnarray}
$a_{5}$ is again an additive background correction. Though the parameter $a_{1}$ is considerably smaller than its Pb counterpart [Eq.~(\ref{eq25a})] we observe that the fit parameters $a_{2}$, $a_{3}$ and $a_{5}$ as obtained from the 0D-model [Eq.~(\ref{eq28b})] and the Gaussian counterpart (Table \ref{Table3}) nearly coincide at $T=95$~K. In analogy to La$_{2-x}$Sr$_{x}$CuO$_{4}$ we attribute this agreement to the fact that the temperatures considered here are considerably above $T_{c}\simeq 85$~K. Indeed, the magnetic penetration depth measurements of Osborn {\it et al.}\cite{Osborn} and their analysis,\cite{tscastro} clearly reveal that the temperature window around $T_c$, where thermal fluctuations dominate, is roughly 1~K only.

\begin{table*}[t!]
\caption[~]{Fit parameters entering Eq. (\protect\ref{eq20}) for slightly underdoped Bi2212 including the additive background correction $a_{5}$. The amplitude $a_{2}=H_{\mathrm{m0}}$ is obtained from Eq. (\protect\ref{eq21}) with $T_{c}=85$~K.}
\label{Table3}%
\begin{tabular}{ccccc}
\hline\hline
$T$~(K) & $a_{3}$~(A/mT) & $H_{\mathrm{m}}$~(T) & $a_{5}$~(A/m) & $a_{2}=H_{\mathrm{m0}}$~(T) \\ \hline
90 & 882.09 & 8.89 & -35.7 & 37.18 \\
95 & 945.60 & 14.93 & -6.79 & 44.77 \\
100 & 788.35 & 19.64 & +1.11 & 48.72 \\
110 & 322.64 & 24.5 & +1.07 & 48.25 \\ \hline\hline
\end{tabular}%
\end{table*}

Using $a_{2}=45$~T we obtain with Eqs.~(\ref{eq12}) and (\ref{eq15})
\begin{equation}
R\xi _{0}=\frac{\Phi _{0}}{\pi a_{2}}\simeq 1.46\times 10^{3}~\text{\AA }^{2},
\label{eq29}
\end{equation}
in comparison with $r\xi _{0}\simeq 3.7\times 10^{5}$~\AA $^{2}$ for Pb corresponding to $a_{2}=940$~Oe, and $R\xi _{0}\simeq 2.13\times 10^{3}$~\AA $^{2}$ for La$_{2-x}$Sr$_{x}$CuO$_{4}$ with $T_{c}\simeq 23$~K [Eq. (\ref{eq27})]. To estimate the radius $R$ of the homogeneous domains we invoke $R/\xi_{0}\simeq 15$,\cite{loram} entering the rounding of the specific heat singularity, to obtain
\begin{equation}
R\simeq 148~\text{\AA }
\label{eq30}
\end{equation}
and $\xi _{0}\simeq 10$~\AA , in comparison with $\xi _{0}\simeq 7$~\AA ,\cite{tscastro} derived from the magnetic penetration depth. Noting that 1~A/m=$10~$erg/(cm$^{3}$T) we obtain for $T=95$~K and $a_{3}=945$~AT/m, $V=1.38\times10^{-18}$~cm$^{3}$ and with $R=148$~\AA\ \ for $d$ the estimate
\begin{equation}
d=\frac{V}{\pi R^{2}}=19~\text{\AA .}
\label{eq31}
\end{equation}
To check the reliability of these estimates, based on $R/\xi _{0}\simeq 15$,\cite{loram}, we invoke Eq. (\ref{eq21a}), yielding with $T=95$~K, $\xi_{0}=R/15$~\AA $=9.9$~\AA\ and $d=19$~\AA , $a_{3}/a_{2}^2=0.49$~A/mT, in reasonable agreement with $a_{3}/a_{2}^{2}=0.47$~A/mT, resulting from Table \ref{Table4}.

\begin{figure}[htb]
%\centering
\includegraphics[width=1.0\linewidth]{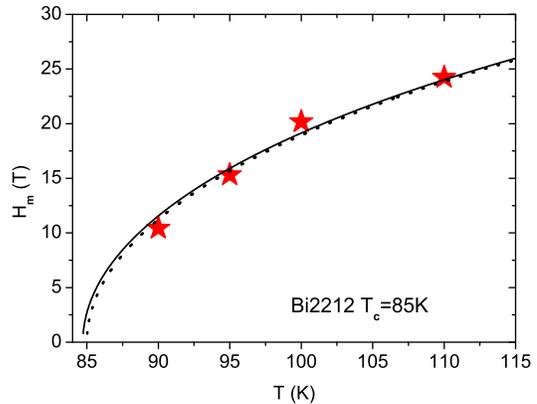}  \vspace{-0.5cm}
\caption{(color online) $H_{\mathrm{m}}$ \textit{vs.} $t$ for Bi2212 with $T_{c}\simeq 85$~K. The solid line is Eq. (\protect\ref{eq19b}) with $a_{1}=311.06$ and $a_{2}=45$~T. The stars denote the $H_{\mathrm{m}}$ values derived from the experimental data as shown in Fig.~\protect\ref{fig4}. The dotted curve is the Gaussian approximation [Eq.~(\protect\ref{eq21})].}
\label{fig7}
\end{figure}

Given the estimates for $a_{1}$ and $a_{2}$ the temperature dependence of $H_{\mathrm{m}}$, the magnetic field where the isothermal magnetization curves adopt a minimum, is readily calculated with Eq.~(\ref{eq19b}) or Eq.~(\ref{eq21}). In Fig.~\ref{fig7}, showing the resulting $H_{\mathrm{m}}\left(t\right) $ we observe excellent agreement with the values derived from the experimental data.

\begin{figure}[htb]
%\centering
\includegraphics[width=1.0\linewidth]{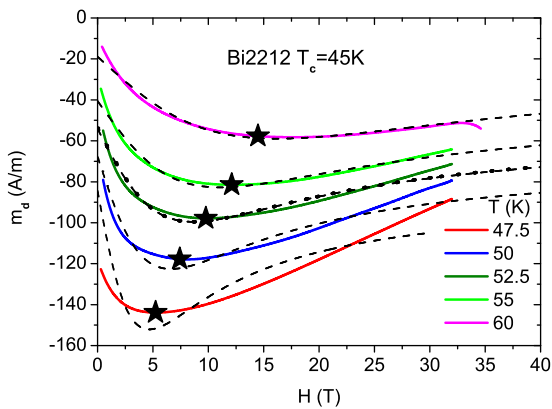}  \vspace{-0.5cm}
\caption{(color online) The solid lines show the isothermal magnetization curves of underdoped Bi2212 with $T_{c}\simeq 45$~K taken from Li \textit{et al}.\protect\cite{ong}. $\bigstar $ indicates the minimum in the isotherms at the respective temperatures. The dashed curves are fits to the Gaussian approximation [Eq. (\protect\ref{eq20})] with the parameters listed in Table \protect\ref{Table4} and the dotted one to the 0D-model [Eq.~(\protect\ref{eq13})] with the parameters given in Eq.~(\protect\ref{eq32}). In both models we included the additive background correction $a_{5}$.}
\label{fig8}
\end{figure}

To explore the effects of doping we consider the isothermal magnetization data of Li \textit{et al}.\cite{ong} for Bi2212 with $T_{c}=45$~K shown in Fig. \ref{fig8}. We include fits of the Gaussian model [Eq. (\ref{eq20})] yielding the parameters listed in Table \ref{Table4} and a fit to the 0D-model [Eq.~(\ref{eq13})] yielding at $T=52.5$~K,
\begin{eqnarray}
a_{1} &=&100.55,\nonumber\\
a_{2} &=& 21.93~\text{T},  \nonumber \\
a_{3} &=&395.01~\text{AT/m},\nonumber\\
a_{5} &=&-53.76~\text{A/m.}
\label{eq32}
\end{eqnarray}
$a_{5}$ is again an additive background correction. Though the parameter $a_{1}$ is considerably smaller than in Bi2212 with $T_{c}=85$~K [Eq. (\ref{eq28b})] we observe that the fit parameters $a_{2}$, $a_{3}$ and $a_{5}$ as obtained from the 0D-model [Eq.~(\ref{eq25a})] and the Gaussian counterpart (Table \ref{Table4}) are close at $T=52.5$~K.

\begin{table*}[t!]
\caption[~]{Fit parameters entering the Gaussian approximation [Eq.~(\protect\ref{eq20})] for heavily underdoped Bi2212 including the additive background correction $a_{5}$. The amplitude $a_{2}=H_{\mathrm{m0}}$ is obtained from Eq.~(\protect\ref{eq21}) with $T_{c}=45$~K.}
\label{Table4}%
\begin{tabular}{ccccc}
\hline\hline
$T$~(K) & $a_{3}$~(A/mT) & $H_{\mathrm{m}}$~(T) & $a_{5}$~(A/m) & $a_{2}=H_{\mathrm{m0}}$~(T) \\ \hline
47.5 & 315.68 & 4.66 & -84.35 & 20.04 \\
50 & 362.76 & 6.61 & -67.7 & 20.36 \\
52.5 & 392.59 & 8.65 & -54.11 & 22.03 \\
55 & 467.32 & 11.07 & -40.5 & 24.17 \\
60 & 548.43 & 16.19 & -18.89 & 30.18 \\ \hline\hline
\end{tabular}
\end{table*}

\begin{figure}[htb]
%\centering
\includegraphics[width=1.0\linewidth]{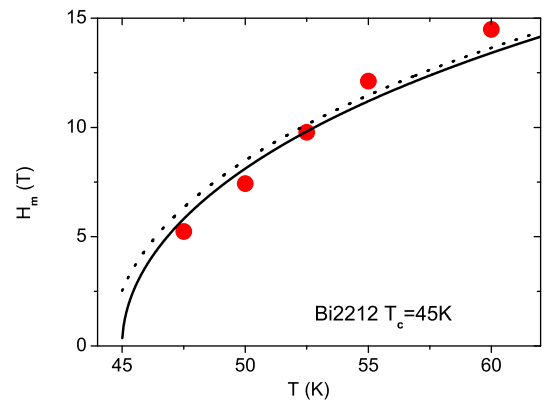}  \vspace{-0.5cm}
\caption{(color online) $H_{\mathrm{m}}$ \textit{vs.} $T$ for Bi2212 with $T_{c}\simeq 45$~K derived from the isothermal magnetization data Li \textit{et al}.\protect\cite{ong}. The dotted line is Eq. (\protect\ref{eq19b}) with $a_{1}=100.55$ and $a_{2}=25$~T. The solid line is the Gaussian approximation [Eq.~(\protect\ref{eq19b})]. }
\label{fig9}
\end{figure}

For this reason the temperature dependence of $H_{\mathrm{m}}$ depicted in Fig.~\ref{fig9} is reasonably well described by the Gaussian approximation Eq.~(\ref{eq21}) with $a_{2}=25$~T, yielding the estimate
\begin{equation}
R\xi _{0}=\frac{\Phi _{0}}{\pi a_{2}}\simeq 2.6\times 10^{3}~\text{\AA }^{2},
\label{eq33}
\end{equation}
in comparison with $R\xi _{0}\simeq 1.46\times 10^{3}$~\AA $^{2}$ for Bi2212 with $T_{c}\simeq 85$~K.

\begin{figure}[htb]
%\centering
\includegraphics[width=1.0\linewidth]{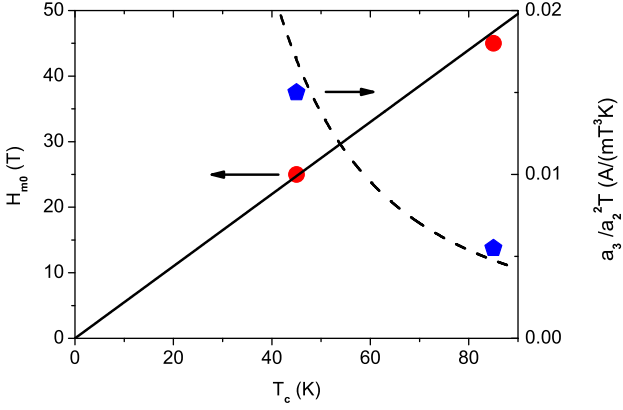}  \vspace{-0.5cm}
\caption{(color online) $a_{2}=H_{\mathrm{m0}}$ \textit{vs.} $T_{c}$ and $a_{3}/(a_{2}^{2}T)$ for underdoped and nearly optimally doped Bi2212. The solid line is $H_{\mathrm{m0}}=a_{2}=0.55T_{c}$\ and the dashed one $a_{3}/(a_{2}^{2}T)=34.43/T_{c}^{2}$. $a_{2}$ and $a_{3}$ are taken from Table \protect\ref{Table3} at $T=95$~K and Table \protect\ref{Table4} at $T=52.5$~K.}
\label{fig10}
\end{figure}

Noting that Bi2212 with $T_{c}=85$~K is close to optimum doping while the sample with $T_{c}=45$~K is underdoped, the amplitude $H_{\mathrm{m0}}$ is expected to exhibit the flow to the quantum phase transition at $T_{c}=0$. Supposing that the doping dependence of $R$, the radius of the homogeneous cylindrical domains is weak $a_{2}=H_{\mathrm{m0}}$ scales according to Eqs.~(\ref{eq12}), (\ref{eq21}) and (\ref{eq25}) as
\begin{equation}
a_{2}=H_{\mathrm{m0}}\propto T_{c}^{1/z},
\label{eq34}
\end{equation}
and $a_{3}/$($a_{2}^{2}T)$ scales according to Eqs.~(\ref{eq21a}) and (\ref{eq25}) as
\begin{equation}
a_{3}/(a_{2}^{2}T)\propto T_{c}^{-2/z}.
\label{eq35}
\end{equation}
In Fig.~\ref{fig10} we plotted our estimates for $a_{2}=H_{\mathrm{m0}}$ and $a_{3}/(a_{2}^{2}T) $ \textit{vs}. $T_{c}$ revealing consistency with the expected flow to the quantum critical point with $z=1$ over an unexpectedly large $T_{c}$ range.

\subsection{Nb$_{0.15}$Si$_{0.85}$}

\begin{figure}[htb]
%\centering
\includegraphics[width=1.0\linewidth]{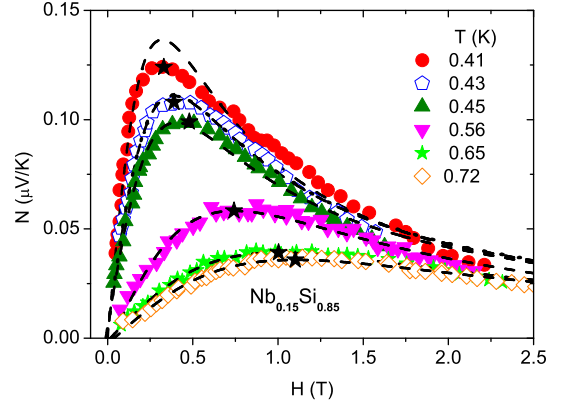}
\vspace{-0.5cm}
\caption{(color online) Isothermal Nernst signal curves of a $350$~\AA\ thick Nb$_{0.15}$Si$_{0.85}$ film with $T_{c}\simeq 0.38$~K taken from Pourret \textit{et al}.\protect\cite{pourret1,pourret,pourret2} The solid lines are fits to the Gaussian approximation [Eq.~(\protect\ref{eq20})] with the parameters listed in Table \protect\ref{Table5}. $\bigstar $ indicates the maximum in the measured isotherms at the respective temperatures.}
\label{fig11}
\end{figure}

The observation of a finite Nernst signal in the normal state of cuprates has revived interest in the study of superconducting fluctuations.\cite{wang,wang2} In conventional superconductors the survival of Cooper pairs above $T_{c}$ has been predominantly examined through the phenomena of paraconductivity\cite{glover} and diamagnetism.\cite{bernardi} To explore the relationship between the Nernst signal and magnetization we consider the data of Pourret \textit{et al}.\cite{pourret1,pourret,pourret2} for a $350$~\AA\ thick Nb$_{0.15}$Si$_{0.85}$ film with $T_{c}\simeq 0.38$~K. In this amorphous superconductor the usual Nernst signal due to normal quasiparticles is negligible.\cite{pourret1,pourret,pourret2} Furthermore, due to the small Hall angle is the Nernst signal simply related to the Peltier coefficient $\alpha _{xy}$ in terms of
\begin{equation}
N=\nu H=\frac{\alpha _{xy}}{\sigma _{xx}}.
\label{eq35a}
\end{equation}
$\nu $ denotes the Nernst coefficient and $\sigma _{xx}$ the conductivity. Above $T_{c}$, as the conductivity changes only weakly with temperature and magnetic field, the evolution of the Peltier coefficient is mainly controlled by the Nernst coefficient.\cite{pourret1,pourret,pourret2} In Fig.~\ref{fig11} we show the isothermal Nernst signal curves for temperatures above $T_{c}$. For comparison we included fits to the Gaussian approximation for the magnetization [Eq.~(\ref{eq20})] with the parameters listed in Table \ref{Table5} and at $T=0.43$~K to the $0$D-model [Eq.~(\ref{eq13})] in terms of the dotted line, yielding the parameters
\begin{eqnarray}
a_{1} &=&355.67,\nonumber\\
a_{2} &=& H_{\rm m0}=1.085~\text{T},  \nonumber \\
a_{3} &=&-0.042~\mu \text{VT/K},\nonumber\\
a_{5} &=& 0.003~\mu \text{V/K}.
\label{eq36}
\end{eqnarray}
With the exception of $T=0.41$~K, which is rather close to $T_{c}\simeq 0.38$~K where fluctuations are expected to contribute, we observe remarkable agreement and a justification of the Gaussian approximation. This agreement also reveals that sufficiently above $T_{c}$ the Nernst signal is proportional to the negative magnetization. Indeed, Fig.~\ref{fig11} clearly reveals that the Nernst signal mirrors the profile of the isothermal magnetization and exhibits the characteristic maximum at $H_{\rm m}$. Because the Gaussian approximation is applicable, the temperature dependence of $H_{\rm m} $ should follow from Eq.~(\ref{eq21}).

\begin{table*}[t!]
\caption[~]{Fit parameters $a_{3}$, $H_{\rm m}$, and $a_{5}$ entering the Gaussian approximation [Eq.~(\protect\ref{eq20})] for the Nb$_{0.15}$Si$_{0.85}$ film. $a_{5}$ is the additive background correction. $H_{\rm m0}$ follows from Eq.~(\protect\ref{eq21}).}
\label{Table5}%
\begin{tabular}{ccccc}
\hline\hline
$T$~(K) & $a_{3}$~($\mu $VT/K) & $H_{m}$~(T) & $a_{5}$~($\mu $V/K) & $a_{2}=H_{m0}$~(T) \\ \hline
0.41 & -0.044 & 0.320 & 0 & 1.161 \\
0.43 & -0.042 & 0.386 & 0.003 & 1.098 \\
0.45 & -0.038 & 0.403 & 0.006 & 0.980 \\
0.56 & -0.047 & 0.742 & 0.003 & 1.192 \\
0.65 & -0.041 & 1.003 & -0.001 & 1.369 \\
0.72 & -0.043 & 1.102 & -0.003 & 1.378 \\ \hline\hline
\end{tabular}%
\end{table*}

\begin{figure}[htb]
%\centering
\includegraphics[width=1.0\linewidth]{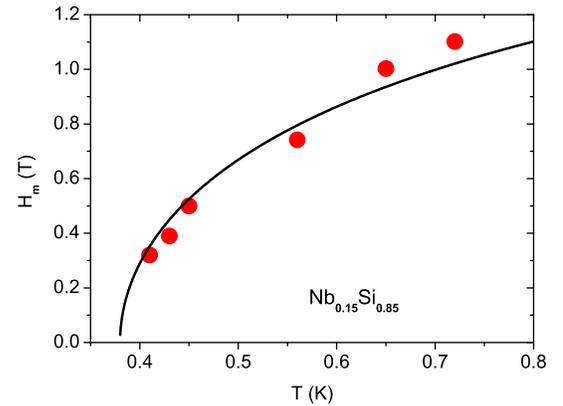}
\vspace{-0.5cm}
\caption{(color online) $H_{m}$ \textit{vs.} $T$ for the Nb$_{0.15}$Si$_{0.85}$ film with $T_{c}\simeq 0.38$~K derived from the data shown in Fig. \protect\ref{fig11}. The solid line is Eq. (\protect\ref{eq19b}) with $H_{\rm m0}=1.28$~T.}
\label{fig12}
\end{figure}

In Fig.~\ref{fig12} we plotted $H_{\rm m}$ \textit{vs}. $T$ and included a fit to Eq.~(\ref{eq21}) yielding $H_{\rm m0}=1.28$~T, in reasonable agreement with the estimates listed in Table \ref{Table5}. It then follows that
\begin{equation}
R\xi _{0}=\frac{\Phi _{0}}{\pi H_{m0}}\simeq 5.2\times 10^{4}~\text{\AA }^{2},
\label{eq37}
\end{equation}
in comparison with $r\xi _{0}\simeq 3.7\times 10^{5}$~\AA $^{2}$ for Pb corresponding to $a_{2}=940$~Oe, $R\xi _{0}\simeq 2.13\times 10^{3}$~\AA $^{2}$ for La$_{2-x}$Sr$_{x}$CuO$_{4}$ with $T_{c}\simeq 23$~K [Eq.~(\ref{eq27})], $R\xi _{0}\simeq 1.46\times 10^{3}$~\AA $^{2}$ in Bi2212 with $T_{c}=85$~K, and $R\xi _{0}\simeq 2.6\times 10^{3}$~\AA $^{2}$ with $T_{c}=45$~K. To estimate the radius $R$ of the homogeneous domains we invoke $\xi _{0}\simeq 130$~\AA,\cite{pourret1,pourret,pourret2} to obtain
\begin{equation}
R\simeq 400~\text{\AA ,}
\label{eq38}
\end{equation}
in comparison with a limiting lateral length of $900$~\AA , obtained from a detailed finite size scaling analysis of the magnetic field dependence of the conductivity in a $125$~\AA\ thick Nb$_{0.15}$Si$_{0.85}$ film.\cite{tstool} This limiting length also implies that the evidence for a magnetic field driven quantum phase transition in this system is constricted by the
resulting finite size effect.\cite{aubin}

Remarkably, the Gaussian version of the 0D-model describes the profile of the isothermal Nernst signal above $T_{c}$ and the temperature dependence of $H_{\rm m}$ in terms of $N\propto -m_{\rm d}$ very well. In the low field limit this relationship transforms with Eq.~(\ref{eq20}) to
\begin{equation}
N\propto -m_{d}\propto -\frac{2a_{3}\pi ^{2}\xi ^{2}R^{2}}{\Phi _{0}^{2}}H,
\label{eq39}
\end{equation}
which differs from the Gaussian fluctuation contribution $N\propto H\xi ^{2}$, valid close to $T_{c}$ and in the zero magnetic field limit.\cite{dorsey,huse}

To substantiate the neglect of thermal fluctuations in the 0D-model further we invoke the Ginzburg criterion for a 2D-system, $\left\vert \Delta T\right\vert /T_{c}\simeq \left\vert 2G_{\rm i}\text{ln}\left( G_{\rm i}\right) \right\vert $. $\Delta T$ is the range of temperatures where thermal fluctuations are essential and $G_{\rm i}=\left( e^{2}/23\hbar \right) R_{\rm n}$ is the is the Ginzburg-Levanyuk parameter for a dirty film with normal state resistance $R_{\rm n}$.\cite{larkin} With $R_{\rm n}=0.3$~k$\Omega $\cite{pourret,pourret2} and $T_{c}=0.38$~K we obtain $\left\vert \Delta T\right\vert \approx 0.01$~K. As a result the Nernst signal curves shown in Fig.~\ref{fig11} were taken outside the critical regime where fluctuations dominate.

\section{Summary and discussion}

\label{sum}

\begin{figure}[htb]
%\centering
\includegraphics[width=1.0\linewidth]{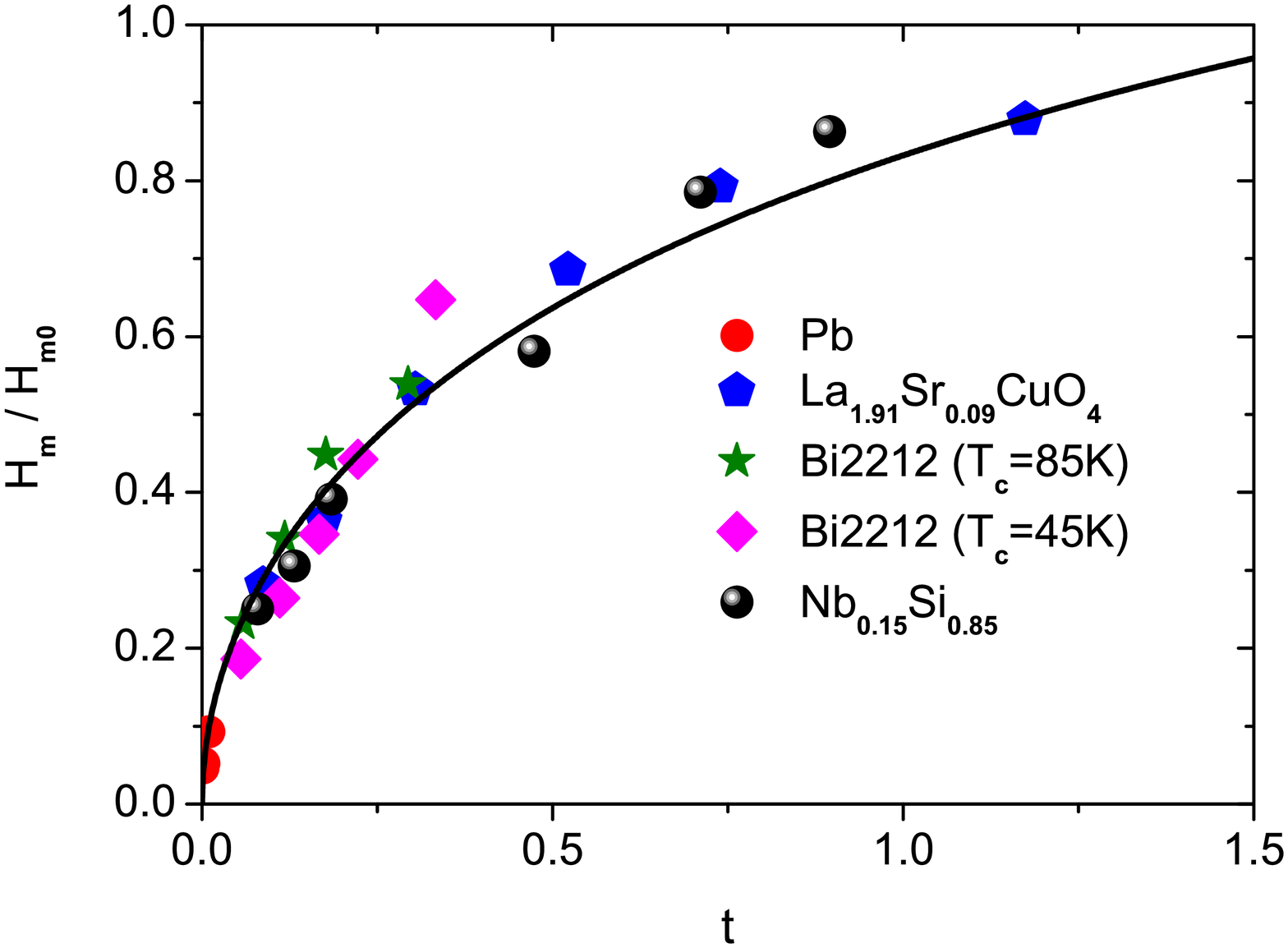}
\vspace{-0.5cm}
\caption{(color online) $H_{\rm m}/H_{\rm m0\text{ }}$\textit{vs}. $t=T/T_{c}-1$ for the data shown in Figs. \protect\ref{fig3}, \protect\ref{fig5}, \protect\ref{fig7}, \protect\ref{fig9}, and \protect\ref{fig12}. The solid curve is $H_{\rm m}/H_{\rm m0\text{ }}=\ln ^{1/2}\left( 1+t\right) $ corresponding to the Gaussian approximation [Eq.~(\protect\ref{eq21})].}
\label{fig13}
\end{figure}

Noting that in cuprate and amorphous conventional superconductors the spatial extent of the homogeneous domains is limited,\cite{tsjs,pan,lang,iguchi,tsbled,loram,tscastro,tstool} we explored the applicability of the 0D-model, neglecting thermal fluctuations, to describe the isothermal magnetization and Nernst signal curves above $T_{c}$. Sufficiently above $T_{c}$ we observed that for both models, the full 0D-model and its Gaussian version, describe the essential features of the curves, including the temperature dependence of the minimum in the magnetization and maximum in the Nernst signal curves at $H_{\rm m}$, rather well. The essential difference between the magnetization curves of the Pb nanoparticles and the bulk cuprates was traced back to the product between the amplitude $\xi _{0}$ of the correlation length and the radius $R$ of the spatial restriction. Indeed, the magnitude of $H_{\rm m}\simeq H_{\rm m0}$ln$^{1/2}(T/T_c)$ is controlled by the amplitude $H_{\rm m0}$. It adopts in the Pb nanoparticles the value $H_{\rm m0}\propto 1/\xi _{0}r\approx 10^{-6}$~\AA $^{-2}$ compared to $1/\xi_{0}R\approx 10^{-5}$ in the $350$~\AA\ thick Nb$_{0.15}$Si$_{0.85}$ film, and $1/\xi _{0}R\approx 10^{-3}$~\AA $^{-2}$ in the cuprates considered here. Indeed, as shown in Fig.~\ref{fig13}, the data for $H_{m}$ vs. $T$, depicted in Figs.~\ref{fig3}, \ref{fig5}, \ref{fig7}, \ref{fig9}, and \ref{fig12} for Pb nanoparticles, La$_{1.91}$Sr$_{0.09}$CuO$_{4}$,
Bi2212 and Nb$_{0.15}$Si$_{0.85}$, tend to fall on a single curve, plotted as $H_{\rm m}/H_{\rm m0\text{ }}$\textit{vs}. $t=T/T_{c}-1$. This curve is even well described by the Gaussian version of the 0D-model [Eq.~(\ref{eq21})]. Thus, for a variety of conventional and hole doped cuprate superconductors it gives a universal perspective on the interplay between diamagnetism, Nernst signal, correlation length, and the limited spatial extent of homogeneity.

Although the assumption of an order parameter which does not depend on the spatial variables fails in the fluctuation dominated temperature window close to $T_{c}$, we established overall agreement between the 0D-model, the isothermal magnetization and the Nernst signal treated as $N\propto -m_{\rm d}$. As a consequence, thermal fluctuations associated with the amplitude and the phase of the order parameter do not contribute significantly in the temperature and magnetic field regimes considered here. The agreement also implies that singlet Cooper pairs in a 0D system subjected to orbital pair breaking are the main source of the observed diamagnetism and Nernst signal in an extended temperature window above $T_{c}$. The monotonic decrease of the magnetization $m_{d}$ and the Nernst signal $N$ with magnetic field $H$ also reveals that there is no particular depairing field. Indeed, $N$ and $m_{\rm d}$ vanish as $\left\vert m_{\rm d}\right\vert \propto \left\vert N\right\vert \propto 2a_{3}/H$ [Eq.~(\ref{eq20})]. Noting that  $N\propto-m_{\rm d}$ also holds for Gaussian fluctuations close to $T_{c}$ and in the zero magnetic field limit,\cite{huse} we have shown that it applies even outside the fluctuation dominated regime.

Clearly, the outlined approach cannot distinguish between intrinsic and extrinsic inhomogeneities, or whether the detected restricted extent of the homogeneous regions reflect an intimate relationship to superconductivity. However, it implies that the reduced dimensionality is not only responsible for the smeared zero field transitions seen in the specific heat,\cite{tsbled,loram}, in the temperature dependence of the magnetic penetration depths,\cite{tsbled,tscastro}, and the resistive transition,\cite{tstool} but also accounts for the observed characteristic minimum in the isothermal magnetization curves and the corresponding maximum in the Nernst signal.

\section{Acknowledgements}
The authors acknowledge stimulating and helpful discussions with H.~Keller. This work was partly supported by the Swiss National Science Foundation.

\end{document}